\documentclass[pra,aps,preprint,superbib,superscriptaddress]{revtex4-2}
\usepackage{amsmath,bm,bbm}
\usepackage{amstext}
\usepackage{epsfig}
\usepackage[table, xcdraw]{xcolor}
\usepackage{subfig}
\usepackage{graphicx}
\usepackage{multirow}
\usepackage{array}
\usepackage{tikz,pgfplots}
\usepackage{MnSymbol}
\usepackage{dsfont}
\usepackage{bbold}
\usepackage{physics}
\usepackage{mathtools}
\usepackage{hyperref}
\usepackage[english]{babel}
\usepackage{ulem}
\usepackage{tikz}
\hypersetup{
    colorlinks=true,
    citecolor=gray,
    filecolor=blue,
    linkcolor=blue,
    urlcolor=red
}
\usepackage{mathrsfs}
\usepackage{xr}
\definecolor{dgreen}{rgb}{0,.5,0}
\definecolor{dred}{rgb}{.7,.0,.0}
\definecolor{olive}{rgb}{0.3, 0.4, .1}
\definecolor{fore}{RGB}{249,242,215}
\definecolor{back}{RGB}{51,51,51}
\definecolor{title}{RGB}{255,0,90}
\definecolor{BlueViolet}{RGB}{138,43,226}
\definecolor{dgreen}{rgb}{0.,0.6,0.}
\definecolor{gold}{rgb}{1.,0.84,0.}
\definecolor{JungleGreen}{cmyk}{0.99,0,0.52,0}
\definecolor{BlueGreen}{cmyk}{0.85,0,0.33,0}
\definecolor{RawSienna}{cmyk}{0,0.72,1,0.45}
\definecolor{Magenta}{cmyk}{0,1,0,0}
\definecolor{wood}{RGB}{139,115,85}
\definecolor{dorange}{RGB}{255,127,0}
\definecolor{dolive}{RGB}{85,107,47}
\definecolor{drg}{RGB}{255,165,0}

\begin{document}
\title{Designing Strain-less Electrode Materials: Computational Analysis of Volume Variations in Li-ion and Na-ion Batteries}
\author{Maxime Maréchal}
\affiliation{ICGM, Université de Montpellier, CNRS, ENSCM, Montpellier, France}
\affiliation{Réseau sur le Stockage Electrochimique de l’Energie (RS2E), 80039 Amiens, France}
\author{Romain Berthelot}
\affiliation{ICGM, Université de Montpellier, CNRS, ENSCM, Montpellier, France}
\affiliation{Réseau sur le Stockage Electrochimique de l’Energie (RS2E), 80039 Amiens, France}
\author{Patrick Rozier}
\affiliation{Réseau sur le Stockage Electrochimique de l’Energie (RS2E), 80039 Amiens, France}
\affiliation{CIRIMAT, Université Toulouse 3 Paul Sabatier, CNRS, 31062 Toulouse, France}
\author{Matthieu Sauban\`ere}
\email{matthieu.saubanere@cnrs.fr}
\affiliation{ICGM, Université de Montpellier, CNRS, ENSCM, Montpellier, France}
\affiliation{Réseau sur le Stockage Electrochimique de l’Energie (RS2E), 80039 Amiens, France}
\affiliation{Univ. Bordeaux, CNRS, LOMA, UMR 5798, F-33400 Talence, France}
\begin{abstract}
  Mechanical degradation in electrode materials during successive electrochemical cycling is critical for battery lifetime and aging properties. A common strategy to mitigate electrode mechanical degradation is to suppress the volume variation induced by Li/Na intercalation/deintercalation, thereby designing strain-less electrodes. In this study, we investigate the electrochemically-induced volume variation in layered and spinel compounds used in Li-ion and Na-ion battery electrode materials through density functional theory computations. Specifically, we propose to decompose the volume variation into electronic, ionic, and structural contributions. Based on this analysis, we suggest methods to separately influence or control each contribution through strategies such as chemical substitution, doping, and polymorphism. Altogether, we conclude that volume variations can be controlled by designing either mechanically hard or compact electrode materials.
\end{abstract}

\maketitle

\section{Introduction}

Continued advancements in lithium-ion batteries have positioned them as the predominant technology for electrical energy storage. 
However, current commercial positive electrode materials undergo significant volume changes during cycling, posing challenges to their long-term performance and stability. For instance, positive electrode materials like layered oxides based on Nickel, Manganese, and Cobalt (NMC's) and the spinel LiMn$_2$O$_4$ experience notable volume alterations in particular when charged to high voltages~\cite{bergNeutronDiffractionStudy1999,caprazElectrochemicalStiffnessChanges2016, kondrakovAnisotropicLatticeStrain2017,jangidRealtimeMonitoringStress2019,ruessInfluenceNCMParticle2020}. The mismatch between the lithiated and delithiated structures, differing in volume and shape, induces an important source of strain leading to mechanical stresses, which are at the origin of electrode material degradation and battery performance deterioration after repeated cycles~\cite{debiasiChemicalStructuralElectronic2019,edgeLithiumIonBattery2021,stallardEffectLithiationShear2022,stallardMechanicalPropertiesCathode2022,devasconcelosChemomechanicsRechargeableBatteries2022,parkZeroStrainCathodesLithiumBased2023}. 
More precisely, these changes affect the general morphology of electrode materials, a well-known cause of capacity loss~\cite{edgeLithiumIonBattery2021}. 
%In a first approximation, the electrochemically-induced strain can be considered as uniform or isotropic and is roughly approximated by the volume difference between the lithiated and delithiated phases renormalized by the number of lithium ions exchanged ~\cite{zhaoDesignPrinciplesZerostrain2022}. 
The primary particles of the cathode, forming secondary particles, undergo repeated attachment and detachment during charging and discharging due to these volume changes~\cite{xuChemomechanicalBehaviorsLayered2018,stallardMechanicalPropertiesCathode2022}. This stress creates cracks inside the particles, leading to a loss of connectivity within the active material. As a consequence, the liquid electrolyte penetrates through these cracks, increasing the cathode electrolyte interface layer. Ultimately, these crystals break down, resulting in significant capacity loss and deterioration of battery performance~\cite{ryuCapacityFadingNiRich2018,ryuCapacityFadingMechanisms2021}.
 These electrochemically-induced strains also pose threats to the separator and overall battery safety. In  solid-state batteries, issues related to volume changes in positive electrode materials are even more problematic~\cite{koerverCapacityFadeSolidState2017,koerverChemomechanicalExpansionLithium2018,shiCharacterizationMechanicalDegradation2020,doerrerHighEnergyDensity2021,yoonDetrimentalEffectHightemperature2022,kalnausSolidstateBatteriesCritical2023}.  Apart from the formation of cracks and fractures in positive electrode materials, "breathing" electrodes lead to the diminution of the interface between the solid electrolyte and the electrode material, posing risks to the system's integrity.
Consequently, the volume changes of the cathode significantly impact battery performances, affecting long-term stability and discharge capacity, presenting a significant challenge to improve high-energy density batteries. Hence, the ideal scenario involves employing cathode materials that undergo minimal volume changes and electrochemically-induced strain during lithium intercalation/deintercalation.

Over the leading factor that controls the electrochemically-induced strain, the nature of the Transition Metal (TM) is shown to play a critical role. More precisely, volume variations and capacity fading have been related, and are drastically modified when changing the Co, Mn and Ni ratio in NMC materials. In particular, high Ni content leads to significant volume changes and mechanical degradation at high voltage~\cite{nohComparisonStructuralElectrochemical2013,sunControlElectrochemicalProperties2015,yoonReviewHighCapacityLi2015a,ishidzuLatticeVolumeChange2016,ryuCapacityFadingNiRich2018,friedrichEditorsChoiceCapacity2019,yuanStabilizingSurfaceChemical2020,liuRationalDesignMechanically2021,parkHighEnergyCathodesPrecision2021,lvReviewModificationStrategies2021}. The crystal structure type is also expected have a significant impact. For instance, in contrast to layered materials, it has been shown that disordered rocksalts, due to cation mixing which leads to a cubic structure, demonstrate minimal and isotropic volume changes during lithiation and delithiation processes~\cite{leeUnlockingPotentialCationDisordered2014a,yabuuchiOriginStabilizationDestabilization2016b,nakajimaLithiumExcessCationDisorderedRocksaltType2017,zhaoDesignPrinciplesZerostrain2022}. 
Besides tuning the TM and the structure, various strategies have been proposed to achieve suppressed volume expansion and contraction towards zero-strain positive electrodes.  In particular electrode surface engineering or concentration gradient~\cite{choZeroStrainIntercalationCathode2001,kimBifunctionalSurfaceCoating2020,liuRationalDesignMechanically2021,nguyenDualProtectiveMechanism2022} and electrode  material doping~\cite{mariyappanRoleDivalentZn2020,zhangImprovingStructuralCyclic2020,parkHighEnergyCathodesPrecision2021,xinWhatRoleNb2021,kongTailoringCo3dO2p2021,ouEnablingHighEnergy2022} have become popular. 

In order to accelerate the design of strain-less electrodes a quantitative or even qualitative evaluation of the different contributions to the electrochemically-induced strain is needed. To that aim, Zhao {\it et al.} have linked qualitatively the volumic variation with the t$_{2g}$ versus e$_g$ character of the redox orbital and confirm that isotropic structures are more prone to reduce the volumic variations~\cite{zhaoDesignPrinciplesZerostrain2022}. 
In this work, we investigate various factors controlling volume changes upon lithium removal in a family of A$_x$MO$_2$ compounds, where A represents an alkali metal (Li or Na) and M represents a 3{\it d} TM. Two structural frameworks have been considered, namely the 2D layered and the 3D spinel structures.  Firstly, our density functional theory computation shows that high spin (HS) and low spin (LS) TM configurations in the intercalated material drastically affects it's cell volume in agreement with the corresponding tabulated ionic radius values. Additionally, van der Walls interactions play a significant role in decreasing the inter layer spacing and thus the volume of deintercalated layered compounds.  In a second step, we propose to decompose the volume variation  in terms of ionic, electronic, and structural contributions. 
Considering the ionic contribution, we show that for Li based compounds the deinsertion process is fairly compensated by the reminiscent electrostatic repulsion of the surrounding anions in the deintercalated phase. The substitution of Li by Na in layered electrode leads to an increase of about $\sim 3.5$\AA$^3$ of the cell volume variation per exchanged alkali, which corresponds to the sphere volumes difference computed using Li$^+$ and Na$^+$ tabulated ionic radii. The electronic contribution is shown to be governed by the nature of the TM and can be estimated using also tabulated ionic radii of both the TM's oxidized and reduced forms. The structural contribution is shown to be important and related to the elasticity of the material. For isotropic compounds, the structural contribution of the volume variation follows the electronic contribution and can be correlated to the isotropic elastic Bulk modulus $B_0$. For layered materials the structural contribution follows the ionic contribution and is isotropic, i.e. the cell is stretched/contracted along the $c$ parameter in order to minimize elastic energy since the elasticity perpendicular to the TM layer is lower than the elasticity in the plane parallel to the TM layer. Finally, based on our analysis we propose at the end of the manuscript hints to reduce volume variations by designing harder or more compact electrode materials.

\section{Computational details}
% Materials and structure
We have studied layered A$_x$MO$_2$ compounds, where A represents an alkali metal, Li or Na at stochiometry $x = 0$ or $1$ and M represents a 3$d$ TM, ranging from Ti to Ni within the {\it R}-3{\it m} layered crystal structure. Spinel compounds Li$_x$M$_2$O$_4$ within the {\it F}d-3{\it m} symmetry group have also been studied for comparison, using the same 3$d$ TM's and for $x= 0$ and 1.
%VASP
Density functional theory calculations were performed using the Vienna Ab initio Simulation Package (VASP)~\cite{kresseEfficientIterativeSchemes1996,kresseEfficiencyAbinitioTotal1996a,kresseUltrasoftPseudopotentialsProjector1999a}. It uses a plane wave basis set and pseudo potentials using the projector augmented wave (PAW) potentials~\cite{blochlProjectorAugmentedwaveMethod1994a}. Both the kinetic energy cut-off and k-point grid were tested, with the criterion for convergence being an energy variation of less than 1~meV/atom. A plane wave energy cutoff of minimum 520 eV and a well converged $\Gamma$-centered Monkhorst–Pack k-point grid were used in these  calculations~\cite{monkhorstSpecialPointsBrillouinzone1976}.  The energy difference convergence criteria of the global break condition for the electronic self-consistent loop was set to 10$^{-8}$eV, whereas for the energy difference convergence criteria of the break condition for the ionic relaxation loop was set to 10$^{-3}$eV.
%SCAN
Structural relaxations and energy calculations were performed using the Strongly Constrained and Appropriately Normed (SCAN) functional~\cite{sunStronglyConstrainedAppropriately2015,sunAccurateFirstprinciplesStructures2016,perdewSemilocalDensityFunctionals2016,zhangEfficientFirstprinciplesPrediction2018} with the inclusion of van der Waals (vdW) interactions by the means of the revised Vydrov–van Voorhis nonlocal correlation functional (rVV10)~\cite{vydrovNonlocalVanWaals2010} called hereafter SCAN-rvv10 functional. 
% PBE+U+D3
These relaxations were also performed with the (rotationally invariant)  DFT+U approach~\cite{dudarevElectronenergylossSpectraStructural1998a} on top of the  Perdew-Burke-Ernzerhof (PBE) functional~\cite{perdewGeneralizedGradientApproximation1996d}.  On-site Coulomb interaction strength ($U$) has been specified for each TM following the litterature, i.e.  U$_{\rm Ti}$ = 3.3~eV, U$_{\rm V}$ = 2.5~eV, U$_{\rm Cr}$ = 3.9~eV, U$_{\rm Mn}$ = 3.9~eV, U$_{\rm Fe}$ = 4.2~eV, U$_{\rm Co}$ = 3.2~eV, U$_{\rm Ni}$ = 6.0~eV. Finally PBE+U calculations have been performed with and without the inclusion of van der Waals interactions by the means of the DFT-D3 method of Grimme {\it et al.}~\cite{grimmeConsistentAccurateInitio2010}. 
% Magnetism
All of these materials were assumed to be ferromagnetic and both the HS and LS configurations were calculated. To access these configurations the difference between the total number of up and down spin electrons was fixed. The atomic charge distribution was computed using the Voronoi deformation density ~\cite{fonseca2004voronoi}.

\section{Results}

\subsection{Volumes and relation with ionic radii}

\begin{figure}[h]
  \centering
  \includegraphics[width=0.45\columnwidth]{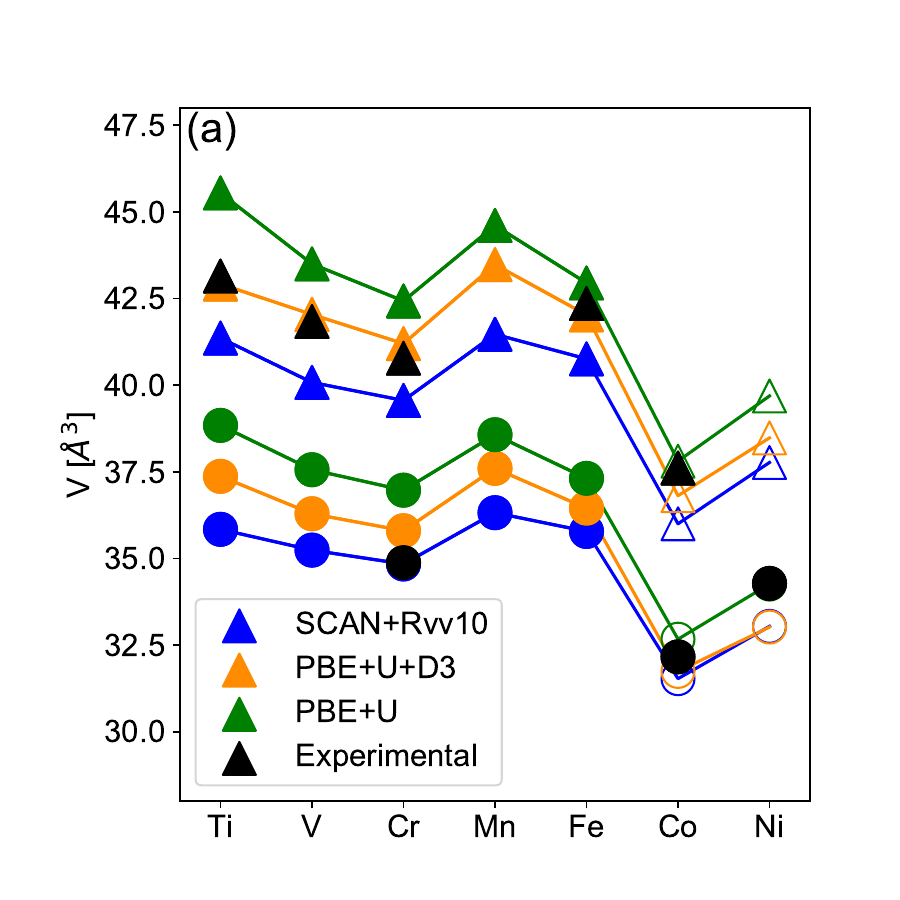}
  \includegraphics[width=0.45\columnwidth]{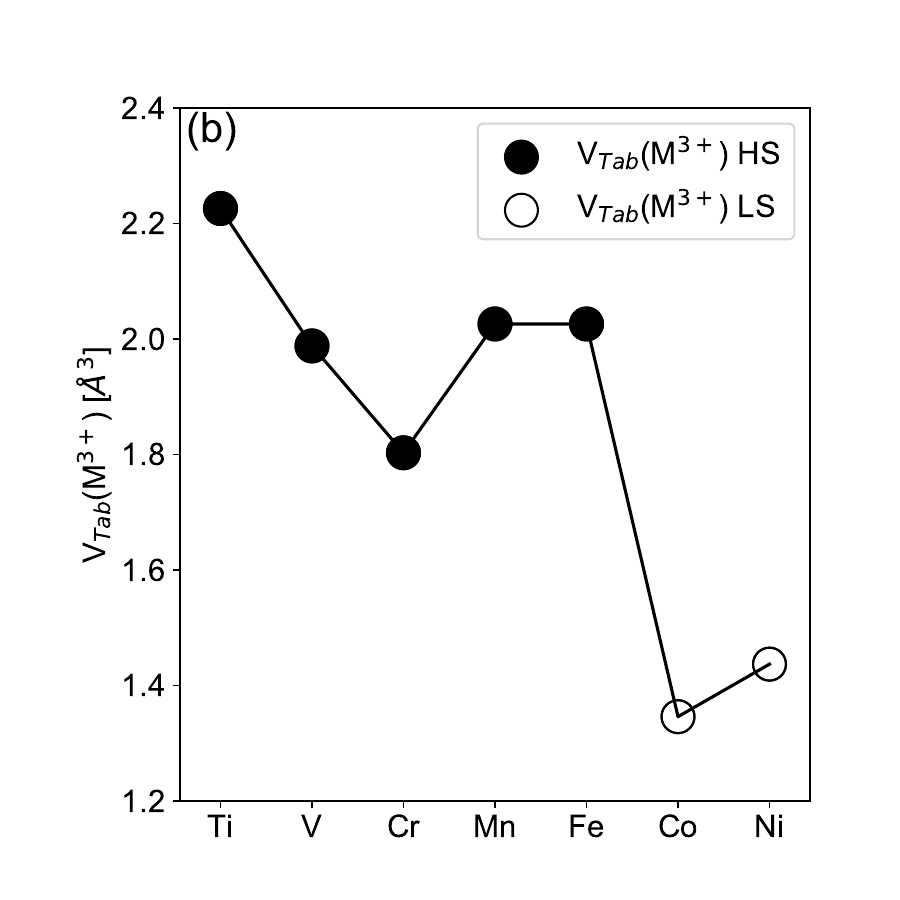}\\
  \includegraphics[width=0.45\columnwidth]{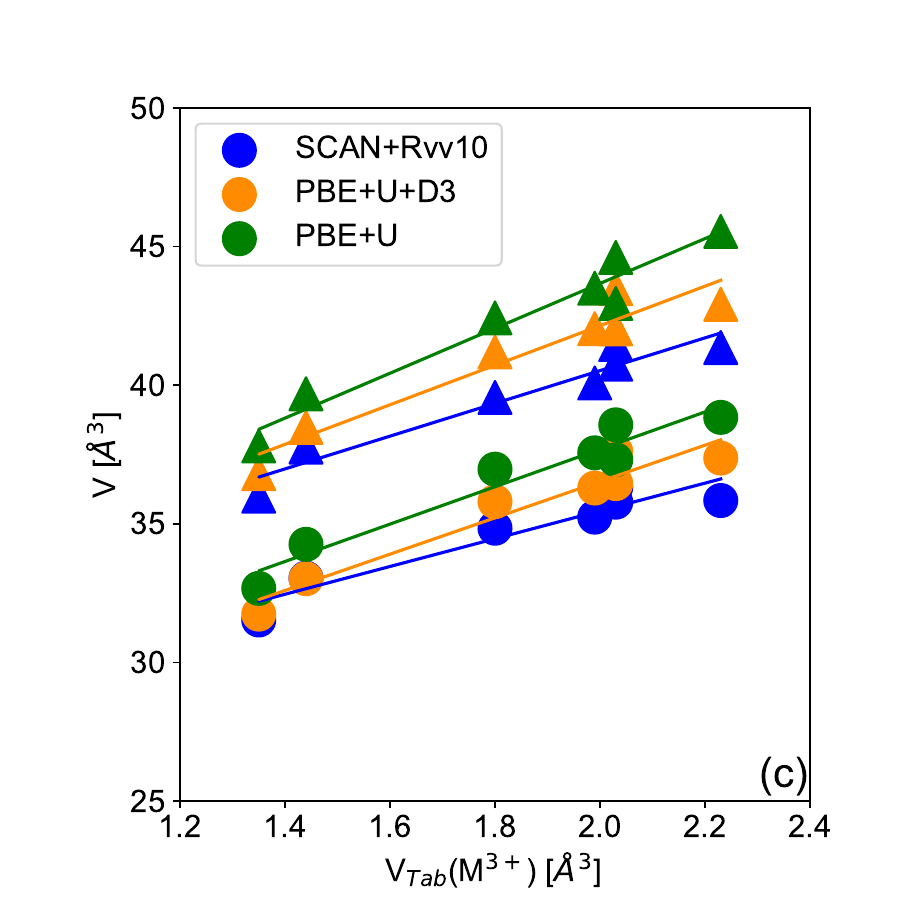}
  \caption{Relaxed volume per MO$_2$ stoichiometry of layered LiMO$_2$ and NaMO$_2$ (a) materials for different 3$d$ transition metals ranging from Ti to Ni. Results are given for the lowest energy spin configuration obtained using different functionals, namey PBE+U (green symbols), PBE+U+D3 (orange symbols) and SCAN-rvv10 (blue symbols) compared with experimental data obtained from the literature (black symbols). Tabulated ionic volume (b) for the LS and HS configurations of the different 3$d$ transition metals in their +III state and in an octahedral environment. In Fig (c) is shown the correlation between the relaxed volume of LiMO$_2$ (circles) and NaMO$_2$ (triangles) with the ionic volume computed using tabulated ionic radii (see text for details). Lines are guides for the eyes.}
  \label{fig:Volume_AMO}
\end{figure}
\begin{figure}[h]
  \centering
  \includegraphics[width=0.45\columnwidth]{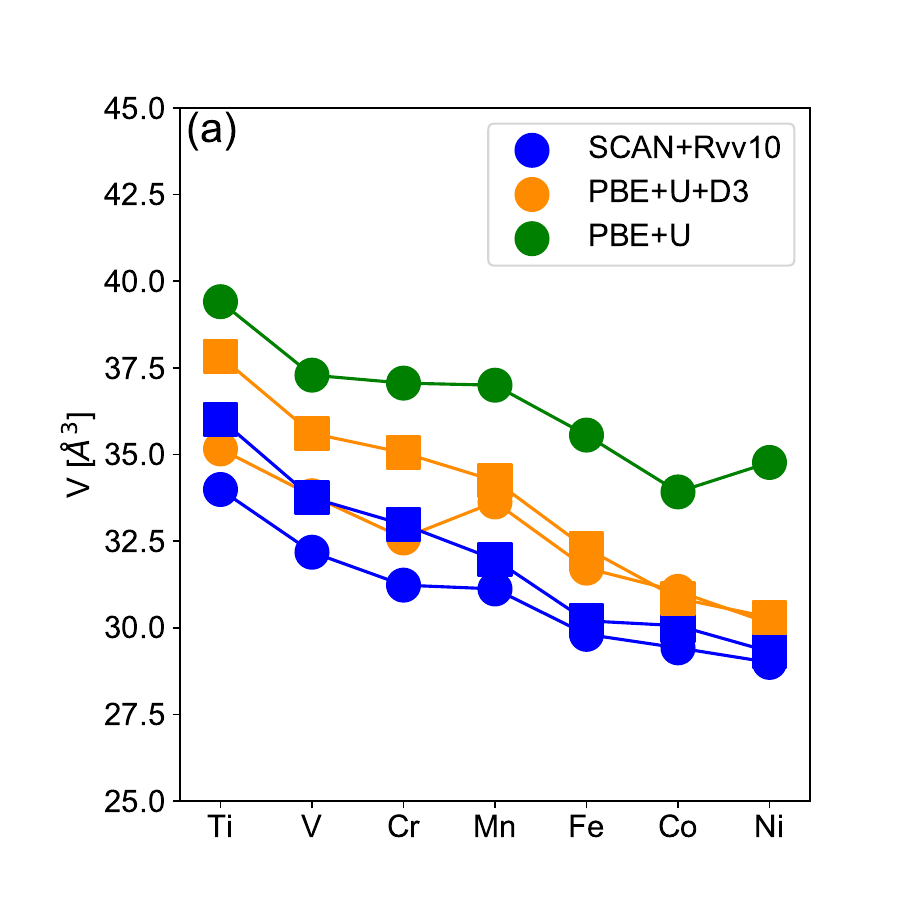}
  \includegraphics[width=0.45\columnwidth]{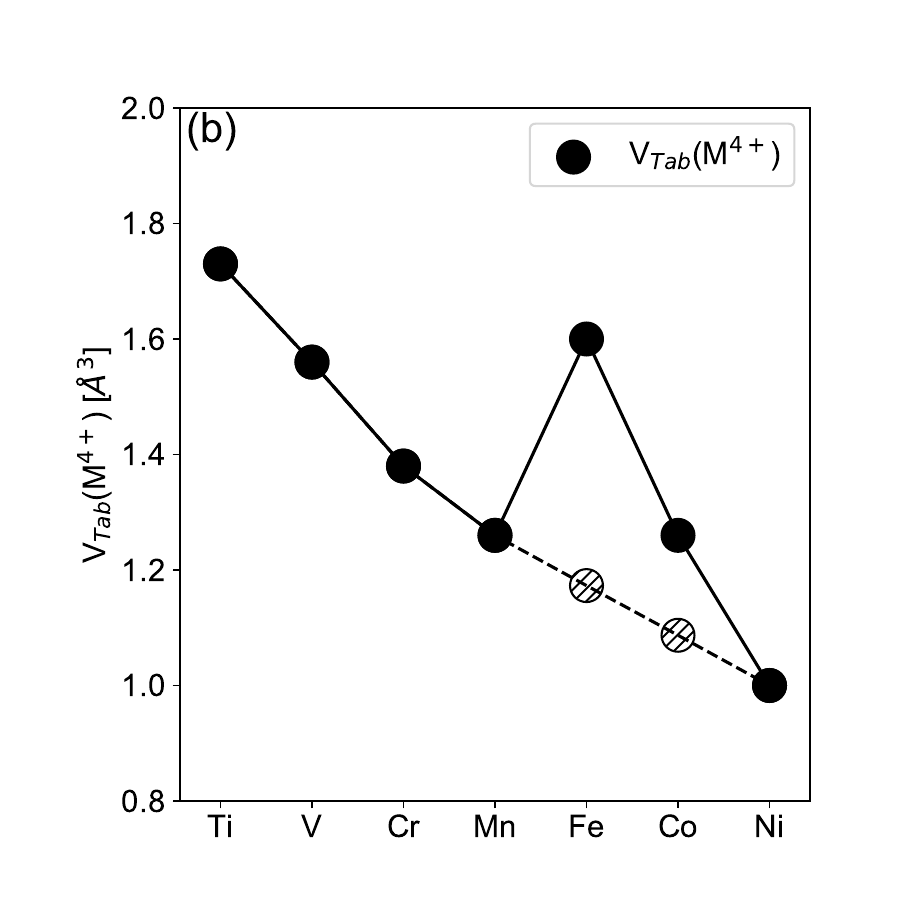} \\
  \includegraphics[width=0.45\columnwidth]{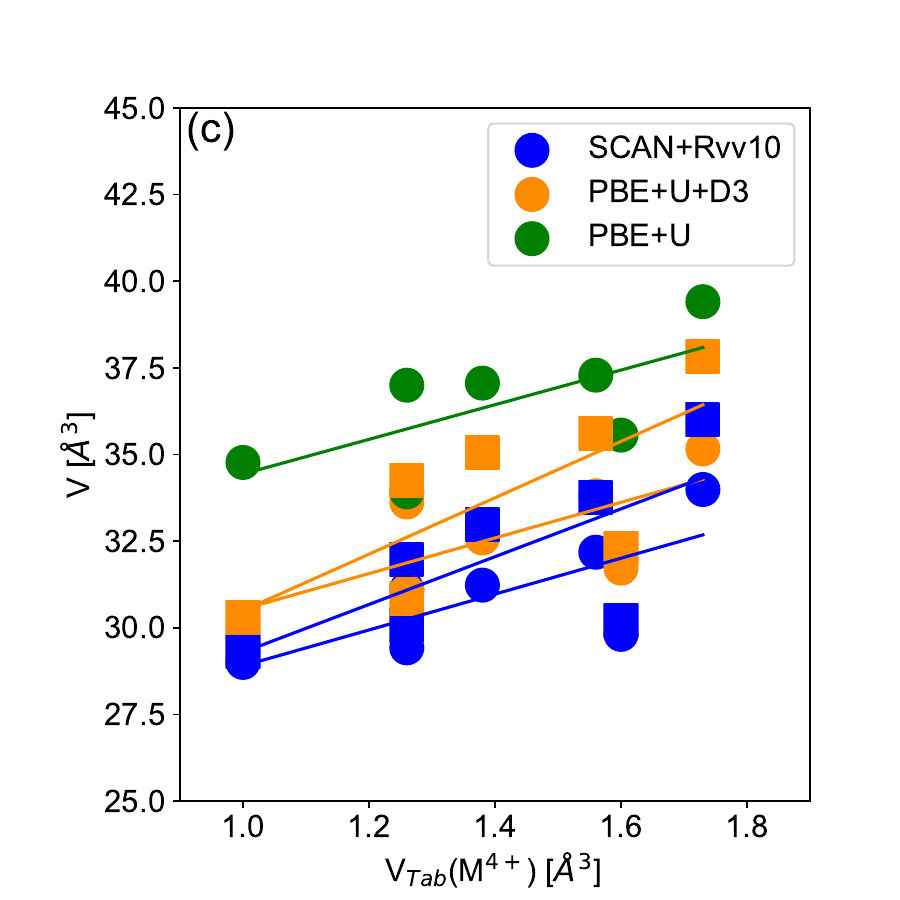}
  \includegraphics[width=0.45\columnwidth]{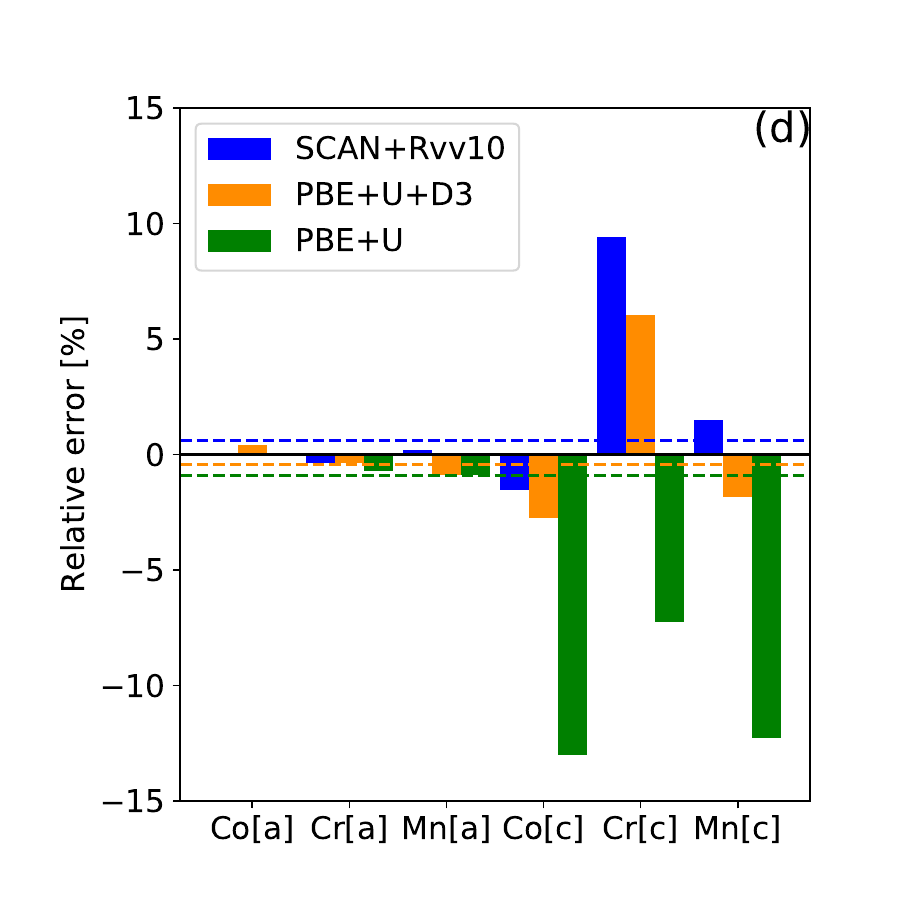}
  \caption{(a) Relaxed volume per MO$_2$ unit of deintercalated MO$_2$ for the layered (circle) and spinel (square) polymorphs for different 3$d$ transition metals ranging from Ti to Ni. Results are given for the lower energy spin configuration obtained using different functionals, namely PBE+U (green symbol), PBE+U+D3 (orange symbols) and SCAN-rvv10 (blue symbols) also compared with experimental volumes obtained from the litterature. (b) Tabulated ionic values of the TM's in their +IV state in an octahedral environement. Hatched values for Fe and Co are calculated from the interpolation between Mn and Ni. In Fig (c) is shown the correlation between the relaxed volume of MO$_2$ and and the ionic volume computed with tabulated ionic radii. In (a) and (b) Lines are guide for the eyes and results are given for the LS/HS configuration leading to the minimum energy. (d) Relative difference (in $\%$) between relaxed and experimental cell parameters~\cite{amatucci1996coo2, bo2016layeredtorocksalt, bruce1999new, wang2002synthesis} a and c for layered materials. Lines are guides for the eyes.}
  \label{fig:Volume_MO}
\end{figure}

We first focus on the volume of intercalated (deintercalated) compounds in Fig.\ref{fig:Volume_AMO} ( Fig.~\ref{fig:Volume_MO}), respectively, for the considered materials, using different functionals in order to evaluate our theoretical setup. 
As shown in Fig.\ref{fig:Volume_AMO}, the relaxed volumes of LiMO$_2$ and NaMO$_2$ per formula unit exhibit non-monotonous behavior as a function of the TM for all considered functionals.
% Functional benchmark
Considering different functionals, the SCAN-rvv10 compared to the PBE + U + D3 functional leads to a volume compression, while comparing the PBE + U and the PBE + U + D3 functionals lead to a volume dilatation, respectively. 
When compared with experimental data, see supplementary information 1, a fair agreement is obtained for all functionals, with however the PBE+U+D3 functional yielding the best mean errors and deviations.
% Cell volume behavior
The behavior of the cell volume as a function of the TM is influenced by both the nature of the TM and its spin state. Indeed, as shown in Fig.~\ref{fig:Volume_AMO}(c), the cell volume is linearly correlated to the ionic volume $V_{\rm Tab}({\rm M}^{3+})= 4\pi R^3_{\rm Tab}({\rm M}^{3+})/3$ in an octahedral environment, where $R_{\rm Tab}({\rm M}^{n+})$ is the tabulated ionic radius of the metal M in an oxidation state $n+$, regardless of the functional considered. See supplementary information 2 for the considered values of ionic radii.  
% spin
Importantly, as shown in supplementary information 3, the LS versus HS configurations lead to drastically different results for M = Mn, Fe, Co, and Ni. Concerning the relative stability of the HS vs LS configurations, our calculations indicate that for AMnO$_2$ the HS configuration is energetically favored, while ACoO$_2$ and ANiO$_2$ adopt the LS configuration. LS states are associated with smaller ionic radii compared to HS configurations since for the latter more anti-bonding e$_g$ states are filled and are associated to larger TM-O bonds due to the presence of anti-bonding $\sigma$ type TM-O bonds.
% special case AFeO2
AFeO$_2$ is predicted to be in the HS configuration with PBE+U and with PBE+U+D3, a result supported by experimental data showing that AFeO$_2$ compounds are HS in different polymorphs~\cite{tabuchiEffectCationArrangement1998}. While the SCAN-rvv10 functional also predicts NaFeO$_2$ to be HS but LiFeO$_2$ to be LS. Given that the energy difference between the LS and HS configurations is rather low ($\sim$ 10 meV/Formula unit), we have considered for LiFeO$_2$ the results of the HS configuration in the manuscript for a sake of coherence.
% JT
We observe an important local octahedral Jahn-Teller (JT) distortion which originates from electronic degeneracy in the e$_g$ shell of the AMnO$_2$ HS ($d^4$) and ANiO$_2$ LS ($d^7$) configurations. As expected, it lowers the  {\it R}$-3${\it m} symmetry to {\it C}$2/${\it m} in the relaxed structure. A less important JT distortion is observed in AVO$_2$ and ATiO$_2$ which originates from the t$_{2g}$ degeneracy. Note that the JT effect is not supposed to have a huge effect on the cell volume which is supported by the linear relation between the cell volume and the ionic radius. A similar linear relation between the cell volume and ionic volume is obtained considering the spinel structure, using the mean values between the corresponding ionic radius of M in the 3+ and 4+ oxidation states, see supplementary information 4.

In Fig.~\ref{fig:Volume_MO}, volumes of deintercalated  MO$_2$ layered and spinel compounds are shown, with M ranging from Ti to Ni, and as a function of the tabulated ionic volume of M$^{4+}$ for different functionals.
In contrast with intercalated compounds, no ambiguities regarding the spin state emerge from our calculations, FeO$_2$, CoO$_2$, and NiO$_2$ being in the LS configuration.
% Layered
Concerning layered compounds, the volume decays quasi linearly with the electronic filling which is consistent with a consecutive filling of the t$_{2g}$ orbitals from zero to six electrons. Moreover, Fig.~\ref{fig:Volume_MO}(b) shows that similarly for sodiated and lithiated compounds, the volume of the MO$_2$ structure follows the trend of M$^{4+}$ ionic volumes obtained through tabulated ionic radii, with an exception for the Iron and at a less extend for Cobalt compounds. However, the ionic radii provided in the table for Fe$^{4+}$ and Co$^{4+}$ must be approached with caution since the $4+$ oxidation state for Fe and Co are quite exotic, as only a few compounds have been isolated; most of them being short-lived intermediaries in reactions in particular for Fe. The computed Voronoi volumes in MO$_2$ compounds show as expected a quasi linear decay as a function of the electronic filling, see supplementary information 5.
A similar trend for the cell volume as a function of the transition metals ionic radius is observed for spinel compounds. Note that 3D spinel structures are shown to be more compact than the 2D layered structures.
% func accuracy
Concerning the functionals' accuracies, the trend of the volume as a function of the considered transition metal appears to be independent of the considered functional. However, the PBE+U functional predicts a volume that is about $\sim 10\%$ above the result obtained using the PBE+U+D3 and SCAN-rvv10 functionals. This overestimation obtained with the PBE+U functional can be attributed to an overestimation of the cell parameter $c$. Indeed, Fig.\ref{fig:Volume_MO}(c) shows the relative error between experimental and computed $a$ and $c$ lattice parameters for synthesized compounds. While one observes a relatively weak mismatch between experimental and computed values for the $a$ parameter, regardless of the functional, the $c$ parameter is shown to be largely overestimated by the PBE+U functional compared to experimental values. The PBE+U+D3 and SCAN-rvv10 functionals are shown to provide much more reliable results, particularly for layered CoO$_2$ and NiO$_2$. Yet, the large overestimation of the $c$ parameter can be attributed to a poor estimation of the van der Waals interactions, which are better taken into account within the PBE+U+D3 and SCAN+rvv10 functionals. These results highlight the importance of the van der Waals interactions on the structural properties of layered oxides, particularly at low alkali content.
For spinel materials,  the PBE+U+D3 functional is shown to be sensitively more accurate than the SCAN-rvv10 functional.
%that probably linked with Van der Waals interactions that are expected to play a much less important role in the 3D spinel structure than in the 2D layered. 
Altogether, in terms of functional reliability, PBE+U+D3 and SCAN-rvv10 are shown to provide structural parameters consistent with each other and consistent with experimental data when available.  On the contrary, the PBE+U functional fails taking into account van der Walls interactions, which leads to a poor description of inter-layer spacing in layered compounds. For this reason, the PBE+U functional is discarded for the rest of the study. In terms of energy, the PBE+U+D3 appears to be more relevant to predict the right LS vs HS configurations but also the electrochemical potentials, see supplementary information 6.

\subsection{Electrochemically-induced volume variation}

\begin{figure}[t]
  \centering
  \includegraphics[width=0.45\columnwidth]{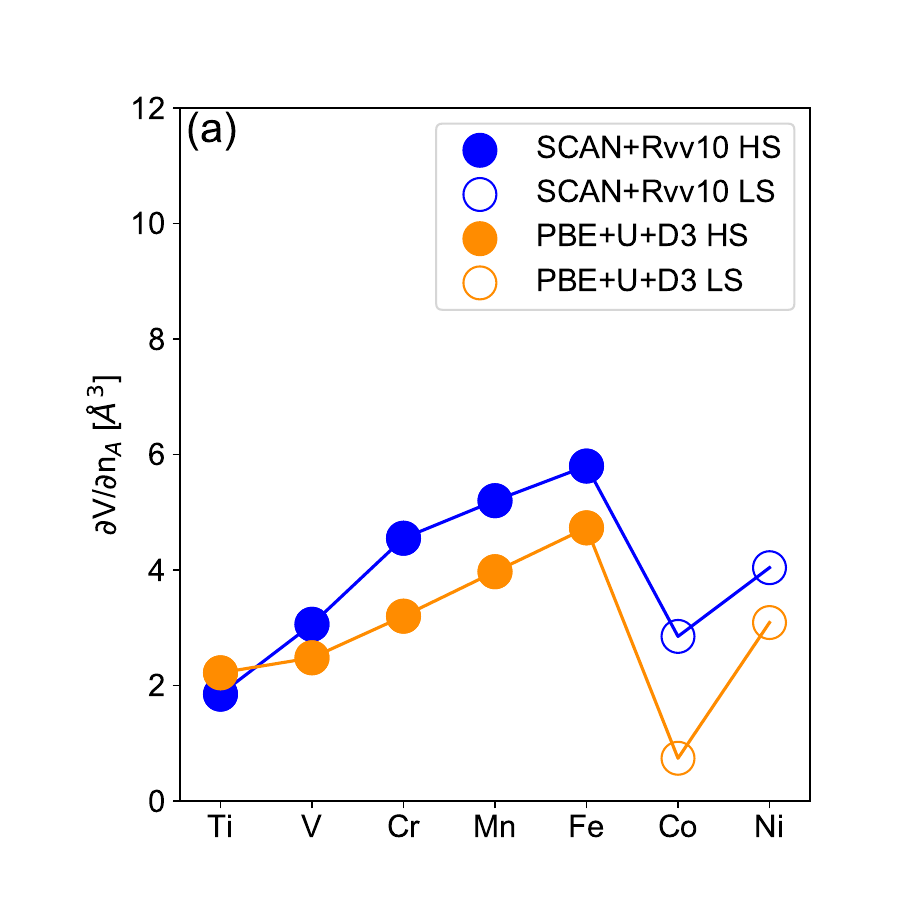}
  \includegraphics[width=0.45\columnwidth]{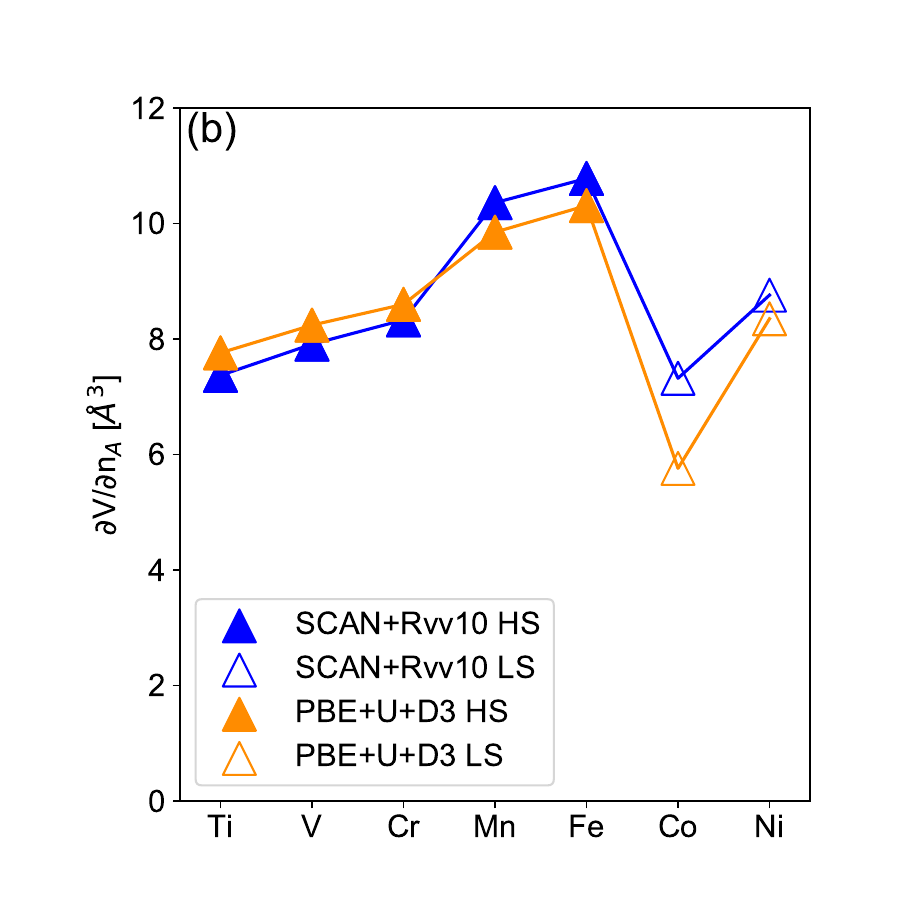}\\
  \includegraphics[width=0.45\columnwidth]{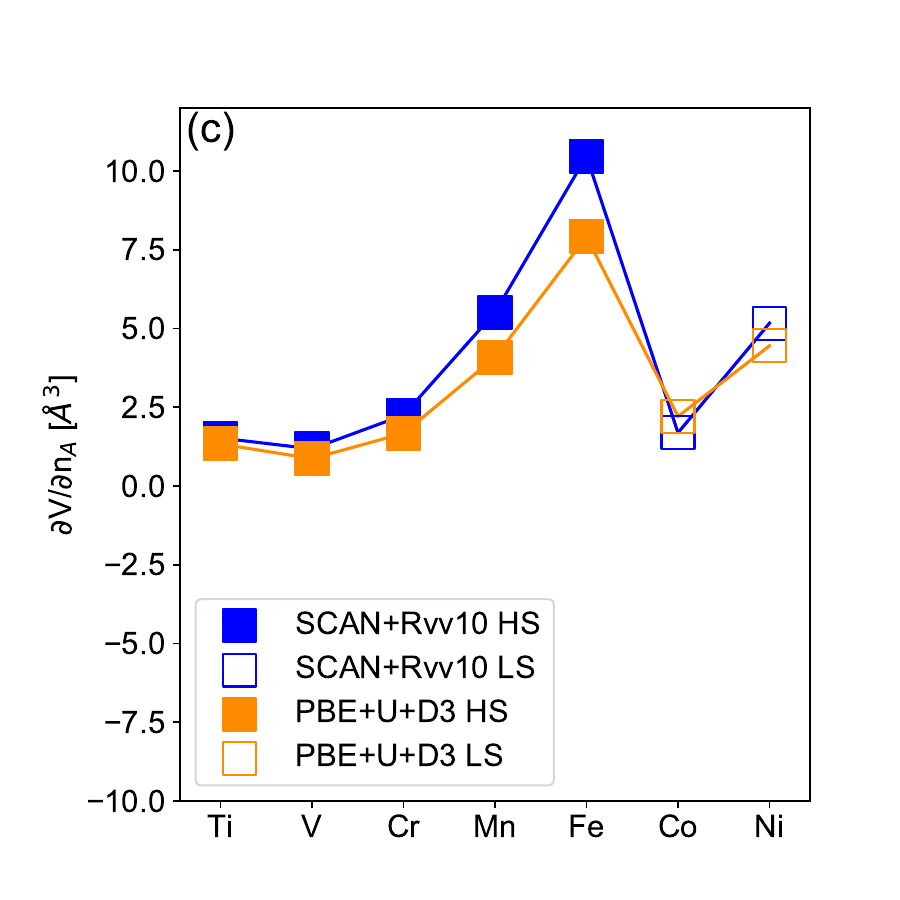}
  \includegraphics[width=0.45\columnwidth]{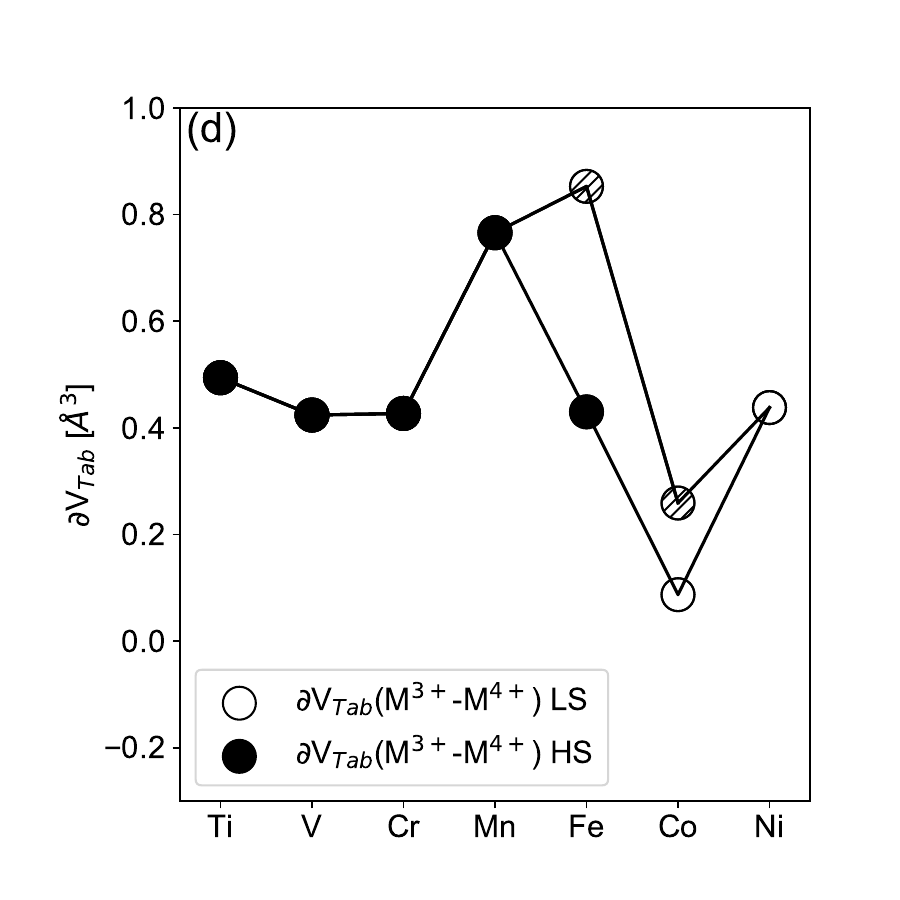}
  \caption{Electrochemically-induced volume variation $\partial V/\partial n_{\rm A}$ of the lithiated (a), sodiated (b) layered structures and for lithiated spinel (c) structure as a function of the transition metal ranging from Ti to Ni. Results are given for the PBE+U+D3 and SCAN-rvv10 functionals. Tabulated ionic volume difference (d) between the +IV and +III (HS/LS) states. For all subfigures, filled (open) symbols highlight cases for witch the intercalated compound is the most stable within the HS (LS)  configuration, respectively. Lines are guides for the eyes.}
  \label{fig:DV}
\end{figure}
\begin{table}
    \centering
    \setlength{\tabcolsep}{0pt} % Reduce the space between columns
    \renewcommand{\arraystretch}{1.2} % Adjust the space between rows
    \begin{tabular}{|c|c|c|c|c|c|c|c|}
        \hline
        \textbf{ LiMeO$_2$ (+III) } & \textbf{ Ti (d$^{1}$) } & \textbf{ V (d$^{2}$) } & \textbf{ Cr (d$^{3}$) } & \textbf{ Mn (d$^{4}$) } & \textbf{ Fe (d$^{5}$) } & \textbf{ Co (d$^{6}$) } & \textbf{ Ni (d$^{7}$) } \\
        \hline
        HS (T2g/Eg) & 
        \cellcolor{green}\raisebox{-0.4\height}{\includegraphics[width=1cm]{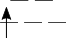}} & 
        \cellcolor{green}\raisebox{-0.4\height}{\includegraphics[width=1cm]{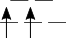}} & 
        \cellcolor{green}\raisebox{-0.4\height}{\includegraphics[width=1cm]{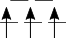}} & 
        \cellcolor{green}\raisebox{-0.3\height}{\includegraphics[width=1cm]{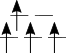}} & 
        \cellcolor{orange}\raisebox{-0.3\height}{\includegraphics[width=1cm]{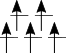}} & 
        \raisebox{-0.3\height}{\includegraphics[width=1cm]{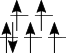}} & 
        \raisebox{-0.3\height}{\includegraphics[width=1cm]{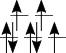}} \\
        \hline
        BS (T2g/Eg) & 
        \raisebox{-0.4\height}{\includegraphics[width=1cm]{1.png}} & 
        \raisebox{-0.4\height}{\includegraphics[width=1cm]{2.png}} & 
        \raisebox{-0.4\height}{\includegraphics[width=1cm]{3.png}} & 
        \raisebox{-0.4\height}{\includegraphics[width=1cm]{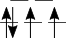}} & 
        \cellcolor{orange}\raisebox{-0.4\height}{\includegraphics[width=1cm]{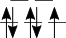}} & 
        \cellcolor{green}\raisebox{-0.4\height}{\includegraphics[width=1cm]{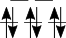}} & 
        \cellcolor{green}\raisebox{-0.3\height}{\includegraphics[width=1cm]{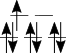}} \\
        \hline
%    \end{tabular}
%\end{table}
%
%\begin{table}
%    \centering
%    \caption{Tableau comparatif des énergies des matériaux MeO2 pour les configurations haut et bas spin SCAN}
%   \setlength{\tabcolsep}{0pt} % Reduce the space between columns
%    \renewcommand{\arraystretch}{1.2} % Adjust the space between rows
%    \begin{tabular}{|c|c|c|c|c|c|c|c|}
        \hline
        \textbf{ MeO$_2$ (+IV) } & \textbf{ Ti (d$^{0}$) } & \textbf{ V (d$^{1}$) } & \textbf{ Cr (d$^{2}$) } & \textbf{ Mn (d$^{3}$) } & \textbf{ Fe (d$^{4}$) } & \textbf{ Co (d$^{5}$) } & \textbf{ Ni (d$^{6}$) } \\
        \hline
        HS (T2g/Eg) & 
        \cellcolor{green}\raisebox{-0.0\height}{\includegraphics[width=1cm]{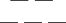}} & 
        \cellcolor{green}\raisebox{-0.4\height}{\includegraphics[width=1cm]{1.png}} & 
        \cellcolor{green}\raisebox{-0.4\height}{\includegraphics[width=1cm]{2.png}} & 
        \cellcolor{green}\raisebox{-0.4\height}{\includegraphics[width=1cm]{3.png}} & 
        \raisebox{-0.3\height}{\includegraphics[width=1cm]{HS_4.png}} & 
        \raisebox{-0.3\height}{\includegraphics[width=1cm]{HS_5.png}} & 
        \raisebox{-0.3\height}{\includegraphics[width=1cm]{HS_6.png}} \\
        \hline
        BS (T2g/Eg) & 
        \raisebox{-0.0\height}{\includegraphics[width=1cm]{0.png}} & 
        \raisebox{-0.4\height}{\includegraphics[width=1cm]{1.png}} & 
        \raisebox{-0.4\height}{\includegraphics[width=1cm]{2.png}} & 
        \raisebox{-0.4\height}{\includegraphics[width=1cm]{3.png}} & 
        \cellcolor{green}\raisebox{-0.4\height}{\includegraphics[width=1cm]{BS_4.png}} & 
        \cellcolor{green}\raisebox{-0.4\height}{\includegraphics[width=1cm]{BS_5.png}} & 
        \cellcolor{green}\raisebox{-0.4\height}{\includegraphics[width=1cm]{BS_6.png}}\\
        \hline
    \end{tabular}
    \caption{Schematic representation of the electronic configurations of each TM with an oxidation degree of 3+ and 4+, also considering the high spin (HS) or low spin (LS) configurations. Cells in green highlight the lowest energy configuration for unambiguous cases (equivalent lowest energy configurations whatever the considered struture (layered {\it vs} spinel), alkali (Li vs. Na) or functional (PBE+U+D3 vs SCAN-rvv10). The case of Fe based compounds is more controversial and is here considered in the HS configuration for layered and spinel compounds (see text for details). %Red circle highligh eg states that are implied in a redox process for Mn, Fe and Ni based materials.)
    } \label{table:Filling}
\end{table}

Given the relaxed intercalated and deintercalated volumes, we can define the electrochemically-induced cell volume variation per alkali A = Li or Na, following Zhao {\it et al}~\cite{zhaoDesignPrinciplesZerostrain2022}.
\begin{equation}
  \frac{\partial V}{\partial n_{\rm A}} = \frac{V_{\rm AHost} - V_{\rm Host}}{n_{A}}
\end{equation}
Where $V_{\rm AHost}$ ($V_{\rm Host}$) is the volume per formula unit of the intercalated (deintercalated) structure, respectively,  and $n_A$ is the number of exchanged alkali per formula unit. In Fig.~\ref{fig:DV} we show $\partial V/\partial n_{\rm A}$  for the layered Li$_x$MO$_2$, Na$_x$MO$_2$ and the spinel Li$_x$Mn$_2$O$_4$ materials using the PBE+U+D3 and the SCAN functionals. Interestingly, the profil of $\partial V/\partial n_{\rm A}$ as a function of the considered TM differs in spinel and layered structures, while it follows the same trend when switching from Li to Na. For layered structures $\partial V/\partial n_{\rm Li}$ increases quasi linearly with the electron filling from 2\AA$^3$ for M = Ti to 6\AA$^3$ for M = Fe. Then it is drastically reduced to 2\AA$^3$ for M = Co before slightly increasing to 2.5\AA$^3$ for M = Ni. The evolution is similar for both functionals and for the sodiated compounds but $\partial V/\partial n_{\rm Na}$ is shifted to higher values following the different ionic radii of Li and Na. For spinel compounds $\partial V/\partial n_{\rm Li}$ shows an overall reduced amplitude compared with layered materials, in particular for M = Ti, V, Cr, Fe and Co. The $\partial V/\partial n_{\rm A}$ trend appears different between spinel and layered materials suggesting that the crystallographic structure plays an important role in governing $\partial V/\partial n_{\rm A}$.
More precisely, while the cell volume of intercalated and deintercalated phases are shown to be qualitatively related to the ionic volume of the TM, $\partial V/\partial n_{\rm A}$ does not follow 
$\Delta V_{\rm Tab} = V_{\rm Tab}({\rm M}^{3+}) - V_{\rm Tab}({\rm M}^{4+})$ for layered electrode materials. 
Consequently, $\partial V/\partial n_{\rm A}$ is controlled by the choice of the TM in spinel compounds, (remember that the value of $\Delta V_{\rm Tab}$ for iron is questionable) but also note that it is strongly influenced by other factors in layered electrodes.  
%
% Discution t2g/eg
In a previous study, Zhao et al. proposed a qualitative argument to rationalize the amplitude of $\partial V/\partial n_{\rm A}$ which is governed by the t$_{2g}$ versus e$_g$ character of electronic states implied in the redox process~\cite{zhaoDesignPrinciplesZerostrain2022}. 
Indeed, the redox associated with e$_g$ states is expected to lead to larger volume variations than for t$_{2g}$ states since the e$_{g}$ (t$_{2g}$) states implies $\sigma$ ($\pi$) type of TM-O bounds, and thus larger (smaller) TM-O distance variation along the redox reaction, respectively. To give rationality to this hypothesis, in Table~\ref{table:Filling}(d) is schematize the electronic filling of the e$_g$/t$_{2g}$ orbitals for the redox active TM's. Following the t$_{2g}$/e$_{g}$ rule would imply that the largest value of $\partial V/\partial n_{\rm A}$ is obtained for Mn (HS), Fe(HS) and Ni (LS) based compounds. For spinel structures the t$_{2g}$/e$_{g}$ rule appears relevant. However, our results suggest that the relation between $\partial V/\partial n_{\rm A}$ and the redox orbital t$_{2g}$/e$_{g}$ character is not straightforward in layered compounds.
For instance, $\partial V/\partial n_{\rm A}$ for the e$_g$ active layered ANiO$_2$ is lower (or equivalent) than for the t$_{2g}$ active AMO$_2$, where M = Ti, Cr and V using the PBE+U+D3 functional. Moreover, the volume variation associated with the deintercalation of an alkali in the HS AFeO$_2$ compounds is associated with a filling change of two e$_g$ electrons, which is not reflected by the volume variation. Finally, $\partial V/\partial n_{\rm Li}$ for the HS LiCoO$_2$ and  HS LiNiO$_2$ compounds, imply a single or double change of the e$_g$ orbital filling, and shows a comparable amplitude for LiCrO$_2$, which is associated with a t$_{2g}$ redox.
%
% -> Need Rationalization -> Decomposition into different contributions
In order to understand deeper the trend of the electrochemically-induced volume variation per alkali, we aim to decompose it in different contributions. From the different possible contributions, one could expect a contribution $\partial V^{\rm Ionic}/\partial n_{\rm A}$ from the additional removal of an alkali ion, proportional to its ionic volume in its environment (Oh versus Td). This contribution is expected to weakly depend on the nature of the TM and on the long range feature of the crystal structure beyond the first coordination sphere. In parallel, we consider an electronic contribution $\partial V^{\rm Elec}/\partial n_{\rm A}$ driven by the removal (addition) of electrons in the redox anti-bonding TM-O  orbitals. This contribution takes into account the contraction  (dilatation) of TM-O bonds upon desintercalation (intercalation), respectively, and is expected to solely depend on the TM's nature and spin configuration. This contribution should be affected by the t$_{2g}$/e$_g$ nature of the redox center.
The decomposition between the different contributions of the electrochemically-induced volume variation can be summarized in the following equation
\begin{equation} 
\frac{\partial V}{\partial n_{\rm A}} = \frac{\partial V^{\rm Ionic}}{\partial n_{\rm A}} + \frac{\partial V^{\rm Elec}}{\partial n_{\rm A}} + \frac{\partial V^{\rm Host}}{\partial n_{\rm A}} \label{eq:DV_decomp}
\end{equation}
where the last term $\partial V^{\rm Host}/\partial n_{\rm A}$ corresponds to a contribution that depends on the Host crystal structure and might modulate ionic and electronic contributions.

\begin{figure}[t]
  \centering
  \includegraphics[width=0.45\columnwidth]{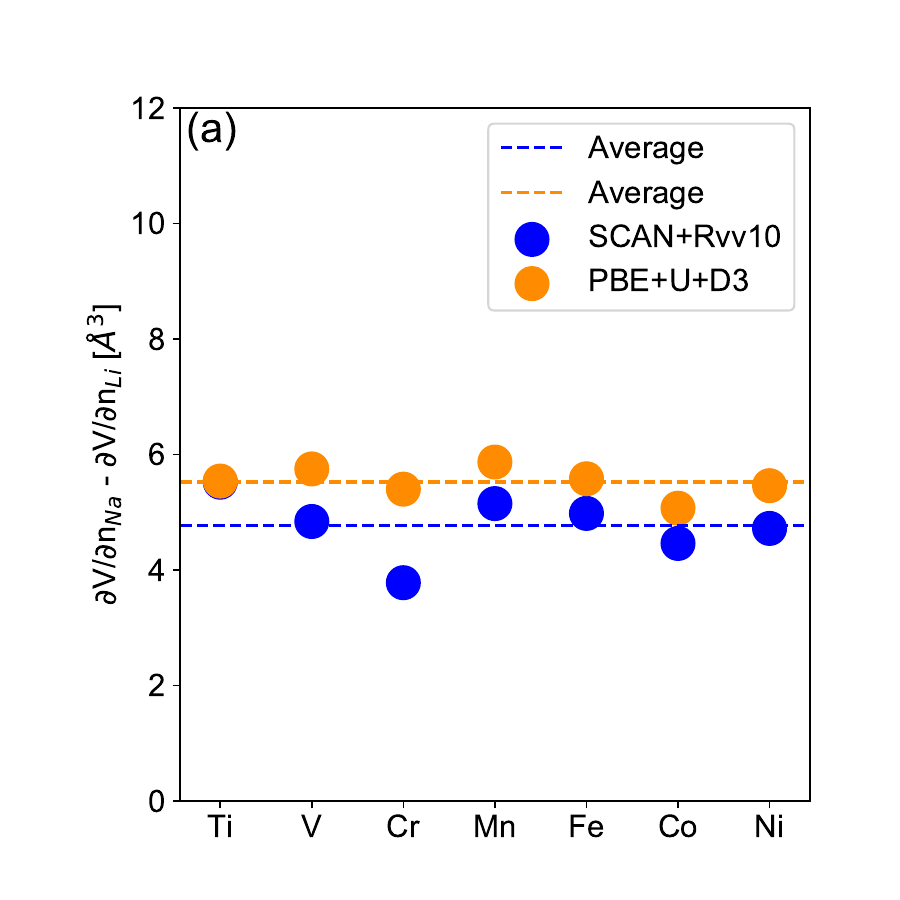}
  \includegraphics[width=0.45\columnwidth]{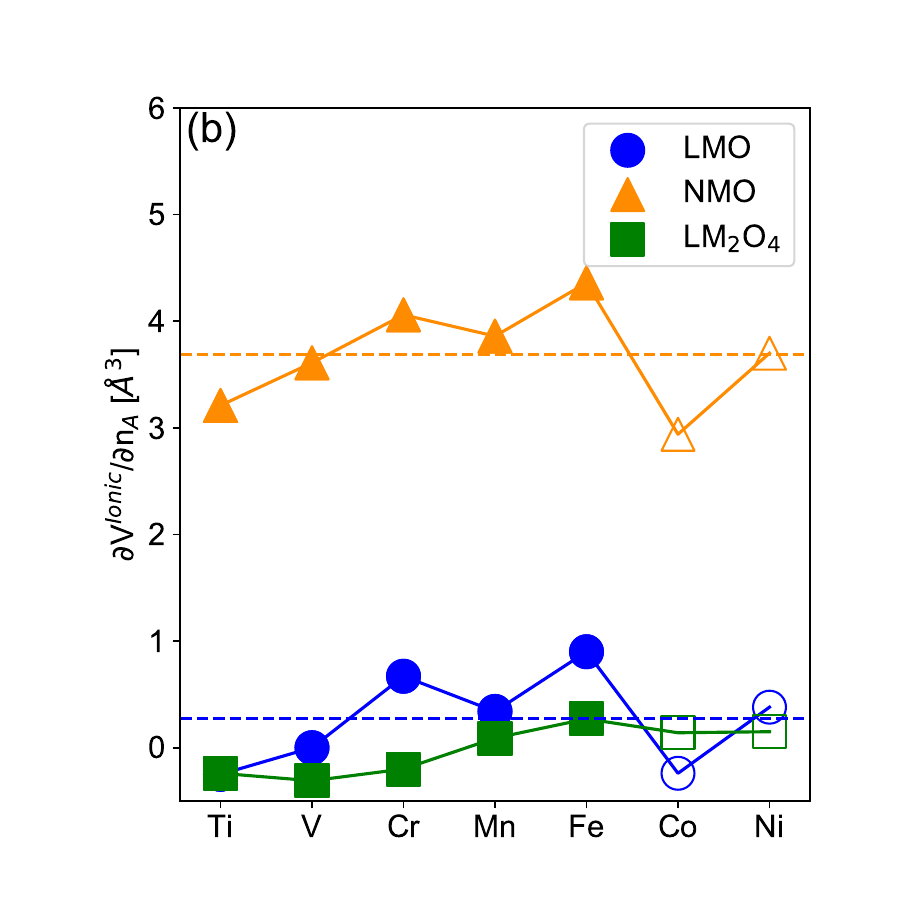}
  \caption{Difference of electrochemically-induced volume variation (a) between the sodiated and lithiated layered materials as a function of the TM for the SCAN-rvv10 (blue circles) and PBE+U+D3 (orange circles) functionals. Ionic contribution (b) of the electrochemically-induced volume variation $\partial V^{\rm Ionic} / \partial n_{\rm A}$ as defined in Eq. (4) as a function of the TM for layered lithiated (blue circles), sodiated (orange triangles) and lithiated spinel (green squares) materials using the PBE+U+D3 functional.}
  \label{fig:DVIonic}
\end{figure}
The amplitude of the $\partial V^{\rm Ionic}/\partial n_{\rm A}$ contribution associated to the addition/removal of an Alkali ion is expected to be proportional to the alkali ionic radius. This assumption is confirmed by our calculations, as shown in Figure~\ref{fig:DVIonic}(a). Indeed, it shows that $\Delta V_{\rm Na\rightarrow Li} =  \partial V/\partial n_{\rm Na} - \partial V/\partial n_{\rm Li}$ is quasi-constant, {\it i.e.} independent of the TM. More precisely, assuming the electronic and structural contributions to be equivalent in iso-structural sodiated and lithiated compounds lead to $\Delta V_{\rm Na\rightarrow Li} \sim \partial V^{\rm Ionic}/\partial n_{\rm Na} -\partial V^{\rm Ionic}/\partial n_{\rm Li}$. Note that the value of $\Delta V_{\rm Na\rightarrow Li} \simeq 5.75$ \AA$^3$ is larger than the difference in atomic volume obtained using tabulated ionic radii $\Delta V_{\rm Tab} = 4\pi \left[R_{\rm Tab}^3({\rm Na}^+) - R_{\rm Tab}^3({\rm Li}^+)\right] / 3 = 3.5$ \AA$^3$, which appears coherent since small cations compress also all other bonds in the structure, an effect that is known as the  chemical pressure. 
To get further into details we compute explicitly $\partial V^{\rm Ionic}/\partial n_{\rm A}$ using the relaxed structure as
\begin{equation}
  \frac{\partial V^{\rm Ionic}}{\partial n_{\rm A}} = \langle V({\rm LiO}_6)\rangle_{\rm AHost} - \langle V(\square \rm{O}_6)\rangle_{\rm Host}
  \label{eq:dV_ionic}
\end{equation}
where $\langle V({\rm AO}_6)\rangle_{\rm AHost}$ and $\langle V(\square \rm{O}_6)\rangle_{\rm Host}$ represent the mean volume of the AO$_6$ octahedra computed in the intercatlated (AHost) compound and the A-vancany $\square$O$_6$ octahedra measured in the deintercalated (Host) structure. Results are presented in Figure~\ref{fig:DVIonic}(b) for lithiated, sodiated layered materials and lithiated spinel materials. $\partial V^{\rm Ionic}/\partial n_{\rm A}$ is shown to be quasi constant for the spinel structure. It follows a trend similar to $\partial V/\partial n_{\rm A}$ with an amplitude of $\sim 1$ (2) \AA$^3$ between the minima for ACoO$_2$ and maxima for AFeO$_2$ for lithiated (sodiated) materials, respectively. Interestingly a mean difference between trend lines for sodiated and lithiated compounds $\Delta V^{\rm Ionic}_{\rm Na\rightarrow Li} = \partial V^{\rm Ionic}/\partial n_{\rm Na} - \partial V^{\rm Ionic}/\partial n_{\rm Li} \simeq 3.5$ \AA$^3$ is found to correspond approximately to the difference between the computed ionic volume and the tabulated ionic radii.
The difference between $\Delta V_{\rm Na\rightarrow Li} $ and $\Delta V^{\rm Ionic}_{\rm Na\rightarrow Li}$ allows to evaluate the effect of chemical pressure to be of about $\Delta V_{\rm Na\rightarrow Li} - \Delta V^{\rm Ionic}_{\rm Na\rightarrow Li} \sim 2.25$ \AA$^3$  by substituting Na for Li.
Another important information from Figure~\ref{fig:DVIonic}(b) concerns the amplitude of $\partial V^{\rm Ionic}/\partial n_{\rm A}$. In particular, for lithiated compounds, $\partial V^{\rm Ionic}/\partial n_{\rm Li}$ shows an amplitude close to zero reflecting the fact that the volume of removed Li$^+$ ions is compensated in the delithiated structure by the electrostatic repulsion between oxygen's in the vacancy octahedra. Similar results are obtained using the SCAN-rvv10 functional, see supplementary information 7. 

\begin{figure}[t]
  \centering
  \includegraphics[width=0.45\columnwidth]{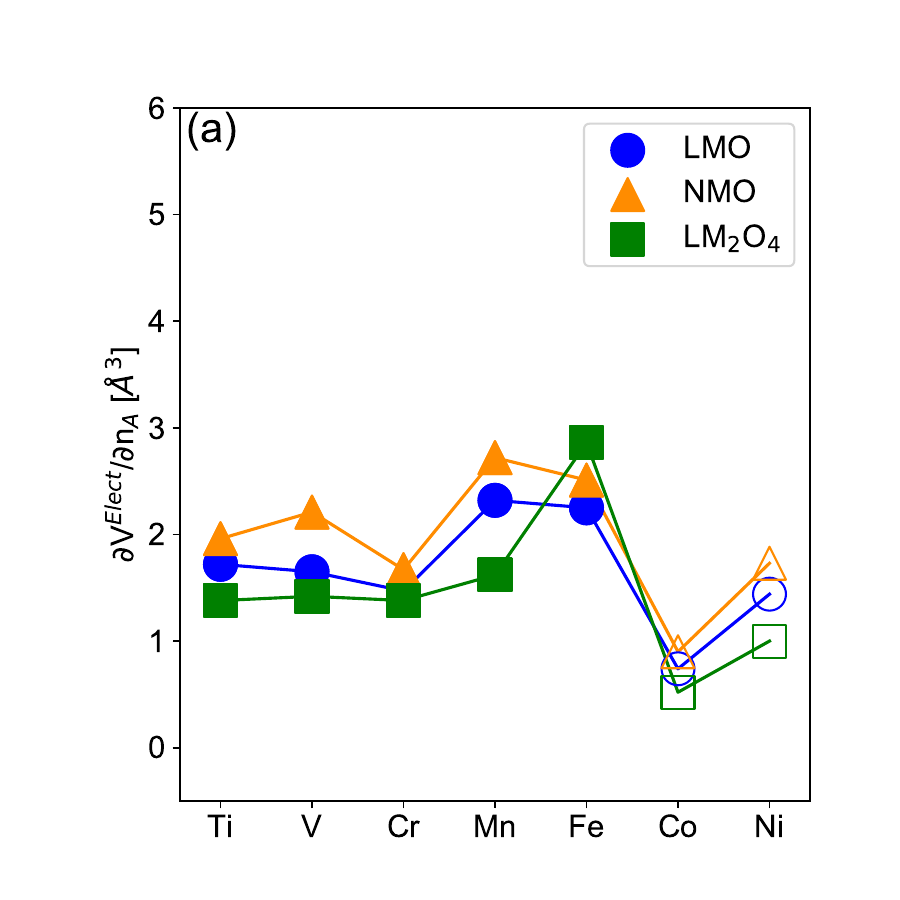}
  \includegraphics[width=0.45\columnwidth]{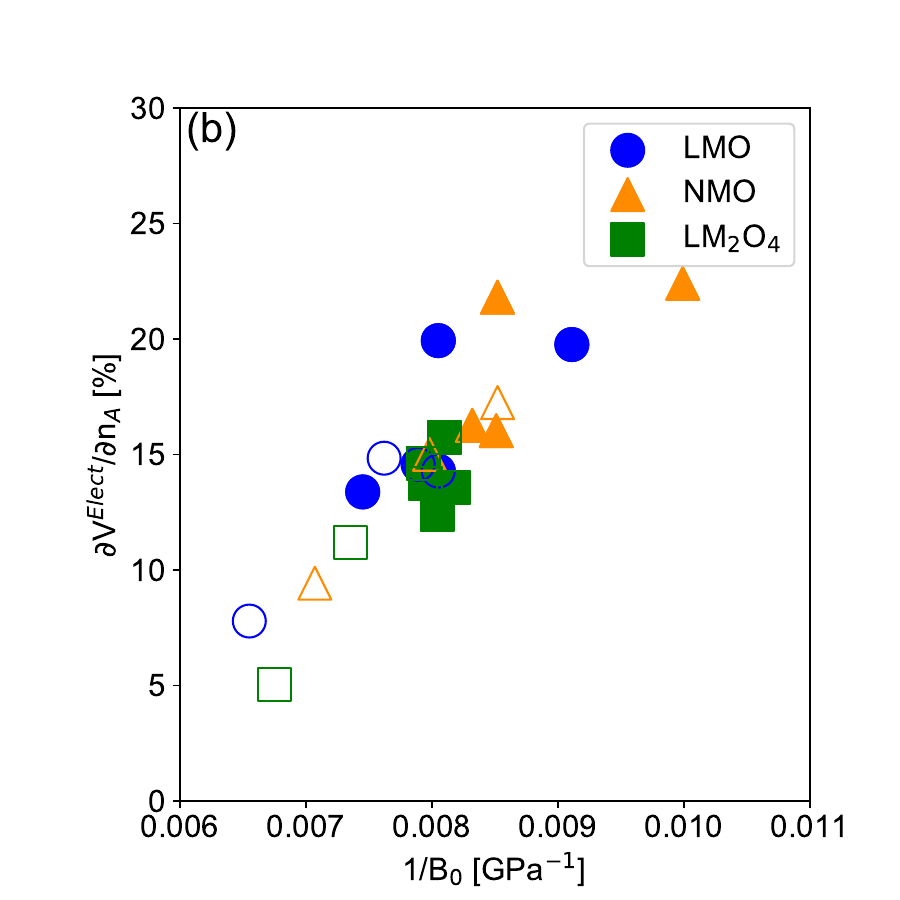} 
  \caption{Electronic contribution (a) of the electrochemically-induced volume variation as defined in Eq.~(\ref{eq:dV_elect}) as a function of the TM for layered lithiated (blue circle), sodiated (orange triangle) and lithiated spinel (green square) materials using the PBE+U+D3 functional. (b) Relation between the electronic part of the total electrochemically-induced strain as a function of the inverse of the Bulk modulus. In all subfigure, filled (Open) symbols corresponds to cases where the intercalated compound is in the HS (LS) configuration, respectively. Lines are guides for the eyes.}
 \label{fig:DVOh}
\end{figure}
We focus now on the induced volume variation associated with the addition/removal of electrons. In order to evaluate $\partial V^{\rm Elec}/\partial n_{\rm A}$ quantitatively, we define the following equation
\begin{equation}
  \frac{\partial V^{\rm Elect}}{\partial n_{\rm A}} = \langle V({\rm MO}_6)\rangle_{\rm AHost} - \langle V(\rm{MO}_6)\rangle_{\rm Host}
  \label{eq:dV_elect}
\end{equation}
where $\langle V({\rm MO}_6)\rangle_{\rm AHost}$ ($\langle V(\rm{MO}_6)\rangle_{\rm Host}$) represents the mean volume of the AO$_6$ octahedra computed in the intercalated (desintercataled) structure, respectively.
In Figure~\ref{fig:DVOh}(a) we show $\partial V^{\rm Elec}/\partial n_{\rm A}$ as a function of the transition metal M for the layered and spinel materials obtained with the PBE+U+D3 functional. Results for $\partial V^{\rm Elec}/\partial n{\rm A}$ appear to be quasi-independent of the considered alkali and the structure type. More precisely, chemical pressure induced by switching from Na to Li does not affect drastically the TM-O bond lengths suggesting that the TMO$_6$ octahedra are the most rigid entities in the structure. This is also supported by the relatively weak influence of the structure on $\partial V^{\rm Elec}/\partial n_{\rm A}$. 
As expected, $\partial V^{\rm Elec}/\partial n_{\rm A}$ is influenced by the t$_{2g}$/e$_g$ character of the redox orbital showing higher amplitudes for Mn, Fe and Ni based materials when Mn and Fe are in the HS configuration. More precisely, $\partial V^{\rm Elec}/\partial n_{\rm A}$  follows the same trend as the variation of the ionic volume computed with the tabulated ionic radii for the transition metal M, $\Delta {V_{\rm Tab}} ({\rm M})$. Importantly, $\Delta {V_{\rm Tab}}$ is tabulated  with respect to the crystal field environment and spin configuration, thus taking into account the change in the t$_{2g}$/e$_{g}$ character of the redox orbital.
% Spinel disproportionation
  Peculiar attention has to be paid to spinel systems. Indeed, some compounds show M$^{3+}$/M$^{4+}$ disproportionation or charge density wave (CDW) in the intercalated phases while in other materials all of the TM's are equivalent, which probably influences the results shown in Figure~\ref{fig:DVOh}(a). More precisely, as illustrated in supplementary information 8, the PBE+U+D3 functional predicts a CDW for all compounds while SCAN-rvv10  predicts CDW only for M = Mn, Co and Ni based LiM$_2$O$_4$ spinel materials. Note that the value of $\partial V^{\rm Elec}/\partial n_{\rm A}$ is not affected by the occurrence of CDW as the volume of M$^{4+}$O$_6$ octahedra in the intercalated structure remains constant upon deintercalation.

  $\partial V/\partial n_{\rm A}$ of the isotropic spinel material  appears to be controlled by its electronic part. This is consistent with the fact that the TMO$_6$ octahedra, being the most rigid part of the structure, should be the limiting factor for the elasticity of the material. Indeed, following Hooke's law, one would expect that the volume change of the host matrix induced by the intercalation is proportional to the elastic modulus of the host matrix which is the case for spinel materials. More precisely, larger (smaller) volume variations $\partial V/\partial n_{\rm A}$ are associated with lower (higher) bulk modulus, or equivalently, with softer (harder) compounds, respectively. On the contrary, for layered materials, the inverse of the Bulk modulus appears to be linearly correlated to $\partial V^{\rm Elec}/\partial n{\rm A}$ rather than $\partial V/\partial n_{\rm A}$ as we show in Figure~\ref{fig:DVOh}. This is consistent with the fact that the bulk modulus or the mechanical hardness of the material is limited by the ionicity of the TM-O bonds, which is governed by the chemical nature of the TM and the electronic filling.  Finally, in an anisotropic structure such as layered materials, the volume variation is also anisotropic and consequently it does not appear as directly related with isotropic elastic moduli such as the bulk modulus.  

\begin{figure}[t]
  \centering
   \includegraphics[width=0.45\columnwidth]{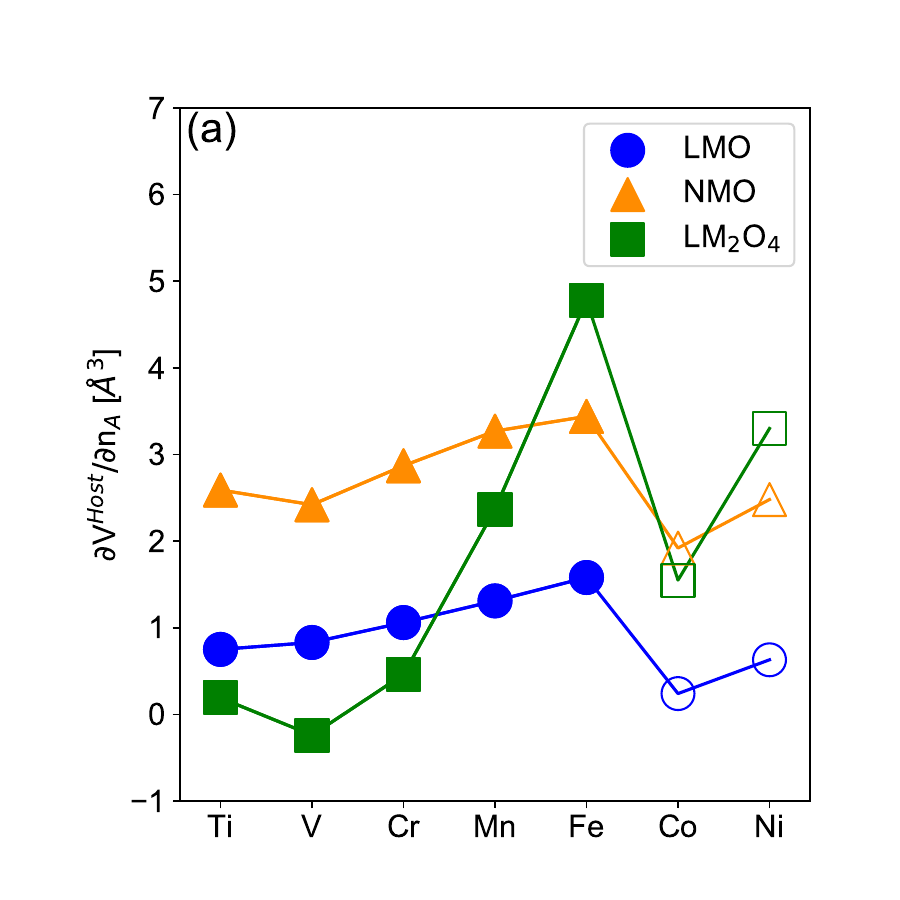}
   \includegraphics[width=0.45\columnwidth]{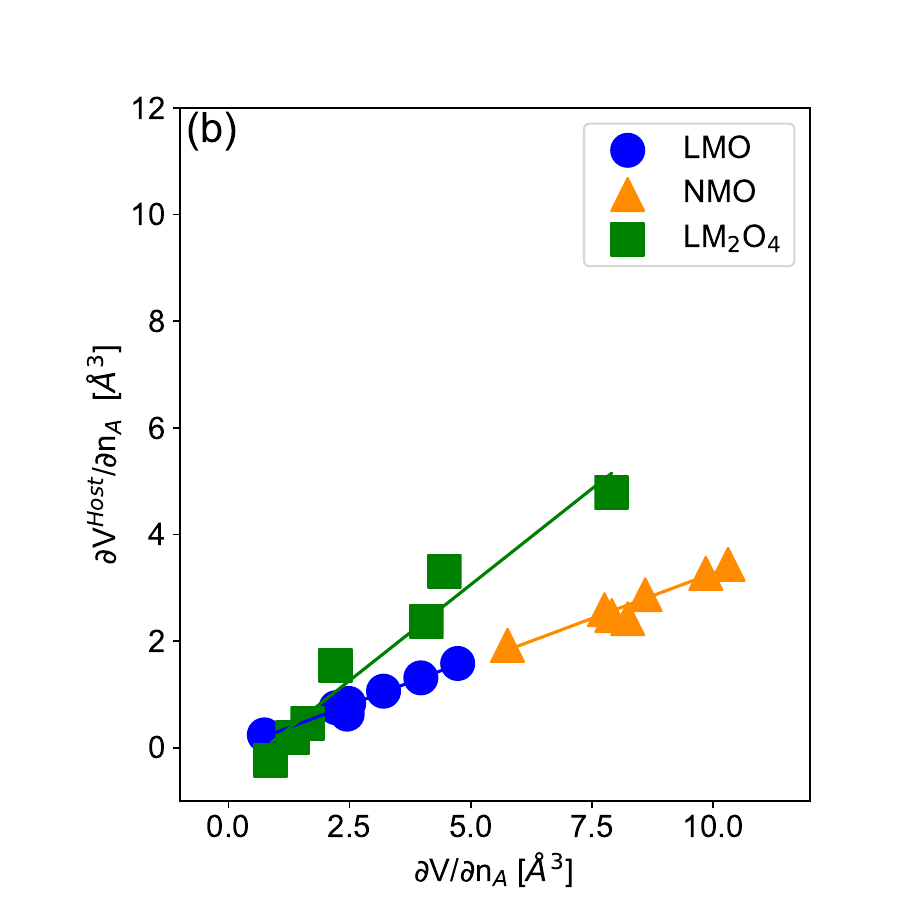}\\
   \includegraphics[width=0.45\columnwidth]{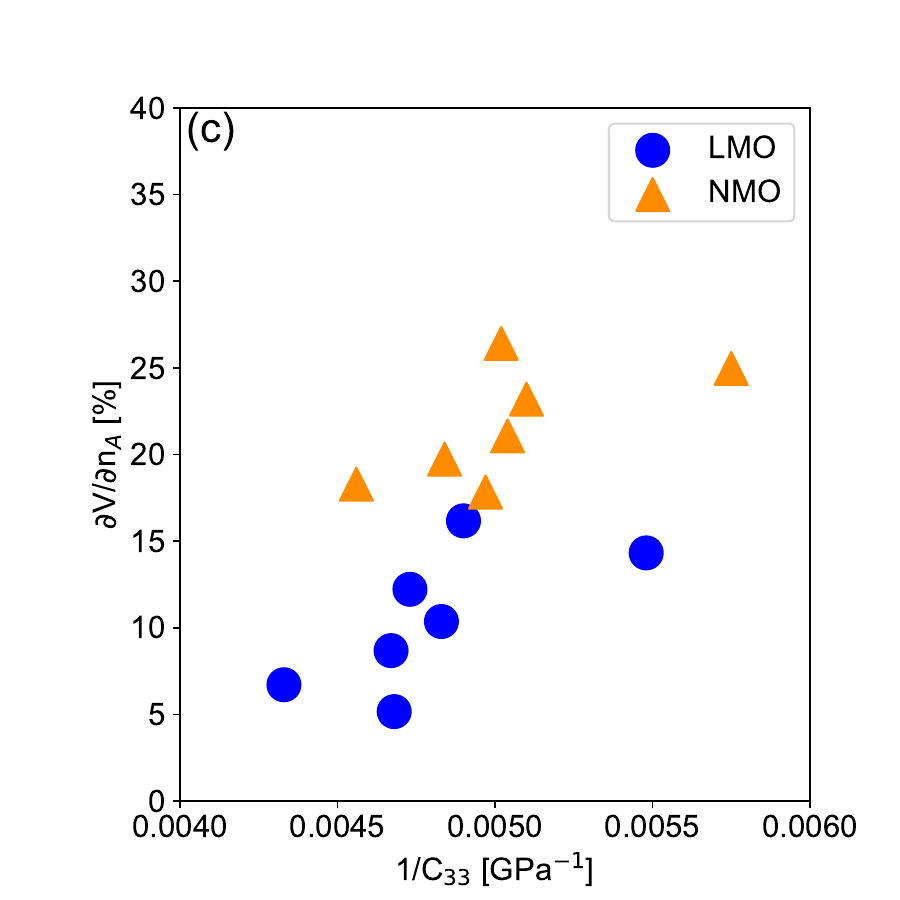}
  \caption{Structural (Host) contribution (a) to the electrochemically-induced volume variation as defined in Eq.~(\ref{eq:dV_elect}) as a function of the TM for layered lithiated (blue circle), sodiated (orange triangle) and lithiated spinel (green square) materials using the PBE+U+D3
functional. (b) Structural (Host) contribution to the electrochemically-induced volume variation as a function of the total electrochemically-induced volume variation. (c) Electrochemically-induced volume variation as a function of the inverse of the elastic contant C$_{33}$. In all subfigures, filled (Open) symbols correspond to cases where the intercalated compound is in the HS (LS) configuration, respectively. Lines are guides for the eyes.}
 \label{fig:DVHost}
\end{figure}

Finally we show in Fig.~(\ref{fig:DVHost}) $\partial V^{\rm Host}/\partial n_{\rm A}$ the contribution to the electrochemically induced volume variation. $\partial V^{\rm Host}/\partial n_{\rm A}$ follows $\partial V^{\rm Ionic}/\partial n_{\rm A}$ for isotropic layered electrodes but $\partial V^{\rm Elec}/\partial n_{\rm A}$ for the isotropic spinels. 
We found a quasi-perfect linear relation between  $\partial V^{\rm Host}/\partial n_{\rm A}$ and $\partial V/\partial n_{\rm A}$ suggesting that $\partial V^{\rm Host}/\partial n_{\rm A}$ results from a compromise between the elasticity along the different TM-O and A-O or $\square$-O bonds. Moreover, in layered material $\partial V^{\rm Host}/\partial n_{\rm A}$ (and consequently $\partial V/\partial n_{\rm A}$) is linearly related to the inverse of the elastic coefficient C$_{33}$ measuring the materials elasticity in the direction perpendicular to the metallic sheets, see Fig.~(\ref{fig:DVHost})(c). As evidenced in other layered materials, this direction is supposed to be soft in particular in deintercalated material as vacancy-O bonds are expected to be soft~\cite{pearceEvidenceAnionicRedox2017e}. Note however that the Jahn-Teller distortion might also influence the elastic coefficient in the different directions, in particular for based Fe and Mn materials, that might explain the non perfect linear trend observed in Fig.~(\ref{fig:DVHost})(c). It contrasts with isotropic electrode materials such as spinels for which the elastic coefficient remains equivalent in all the directions such that $\partial V/\partial n_{\rm A}$ is correlated with the isotropic bulk moduli, as already discussed, dictated by the limiting TM-O bond hardness.
Overall, it suggests that the volume variations are dictated by the elasticity of the material and is consequently strongly influenced by the crystal's structure. 
%More precisely, our results show that the Host contribution to the volume variation results from a competition between the elasticity of the different component/directions. 
Yet the anisotropy of the structure might induce softer (more elastic) directions along which the volume variations are expected to be increased such as for layered materials.
%by which shows that  $\partial V^{\rm Host}/\partial n_{\rm A}$  (and consequently $\partial V^{\rm Ionic}/\partial n_{\rm A}$)  is controlled by elastic coefficients in the direction that corresponds to the direction perpendicular to the layers.  In contrast, as already discussed, for isotropic structures, the elasticity remains equivalent in all directions and the volume variations are dictated by the isotropic Bulk moduli, itself controlled by the hardness of the TM-O bonds.

\section{Discussion and Conclusion}

By decomposing the electrochemically-induced volume variations $\partial V/\partial n_{\rm A}$ between ionic, electronic and structural contributions we have shown that the ionic part depends mainly on the exchanged alkali type, Li vs Na. 
%The ionic contribution can be considered constant regardless of the transition metal and crystal structure, as long as the alkali elements have equivalent octahedral and tetrahedral coordination shells. 
Moreover in the case of lithium the volume loss associated to Li removal is nearly compensated by electrostatic interactions leading to a quasi-nil effect.

The electronic contribution takes into account the t$_{2g}$/e$_g$ character of the redox orbital which plays an important role but does not constitute a sufficient criteria to be predictive. However the electronic contribution is shown to govern $\partial V/\partial n_{\rm A}$ in isotropic materials such as the spinels. Fortunately, it can be quantitatively estimated by using tabulated ionic radii associated to the concerned TM in the relevant environment and oxidation state. This appears to be a reliable descriptor to estimate this contribution at the exception of Fe for which the Fe$^{4+}$ tabulated ionic radius has to be questioned.

Finally, the volume variation associated with the Host reaction to the addition of electrons and alkali ion is shown to result from a compromise between the elasticity associated with the different components. {\it De facto}, it depends strongly on the anisotropic character of the crystal structure.

 Altogether, our results show that the overall volume variations are controlled by the elasticity of the material suggesting that the intercalation is an elastic, thus reversible, process. For isotropic structures the electrochemically induced strain is linearly correlated with the inverse of the Bulk modulus. The case of anisotropic materials is more complex as the elasticity is also anisotropic and the volume variations are consequently expected to be increased in the easiest (softer) direction. Consequently $\partial V/\partial n_{\rm A}$ cannot be directly predicted from tabulated ionic volumes or using the isotropic elastic moduli. In the case of layered material, we show that the volume variations are related to the elastic coefficient in the direction perpendicular to the layer that is believed to be lower than in a direction parallel to the layer. 

In terms of material design, our results suggest that the most efficient way to reduce electrochemically-induced volume variations, besides manipulating the chemical nature of the alkali or transition metal, is to take advantage of the elastic character of intercalation. 
As an evidence, harder materials will experience less volume variations, but they might also be restrained in terms of capacity as the intercalation process might also be limited. The material's harness can be modulated using for instance, chemical substitutions. Note however that to be efficient, the tuning of the hardness has to be performed isotropically such that no softer direction emerges and cancels the desired effect.
Another strategy involves designing compact electrodes, as the volume variation is expected to be proportional to the material's volume in an elastic system. This can be achieved in several ways: by focusing on compact polymorphs (generally 3D rather than 2D), by controlling the spin configurations to be LS, or by chemically doping with "small" ions, such as Al$^{3+}$. In particular, the doping strategy can be especially efficient if, in addition to reducing the volume, it forces the LS configuration and increases the mechanical hardness of the material.

% \begin{figure}[t]
%   \centering
%   (a)\includegraphics[width=0.45\columnwidth]{OO1_XMO.pdf}
%   (b)\includegraphics[width=0.45\columnwidth]{OO2_XMO.pdf}\\
%   (c)\includegraphics[width=0.45\columnwidth]{OO1_MO.pdf}
%   (d)\includegraphics[width=0.45\columnwidth]{OO2_MO.pdf}\\
%   (e)\includegraphics[width=0.45\columnwidth]{OO_XMO.pdf}
%   (f)\includegraphics[width=0.45\columnwidth]{OO_MO.pdf}\\
%   \label{fig:DV_xx}
% \end{figure}

% \begin{figure}[t]
%   \centering
%   (a)\includegraphics[width=0.45\columnwidth]{C33_LMO.pdf}
%   (b)\includegraphics[width=0.45\columnwidth]{C33_NMO.pdf}\\
%   (c)\includegraphics[width=0.45\columnwidth]{C33_LM2O4.pdf}
%   (d)\includegraphics[width=0.45\columnwidth]{C33_MO.pdf}\\
%   (e)\includegraphics[width=0.45\columnwidth]{C33_M2O4.pdf}
%   \label{fig:DV_xx}
% \end{figure}

% \begin{figure}[t]
%   \centering
%   (a)\includegraphics[width=0.45\columnwidth]{C55_LMO.pdf}
%   (b)\includegraphics[width=0.45\columnwidth]{C55_NMO.pdf}\\
%   (c)\includegraphics[width=0.45\columnwidth]{C55_LM2O4.pdf}
%   (d)\includegraphics[width=0.45\columnwidth]{C55_MO.pdf}\\
%   (e)\includegraphics[width=0.45\columnwidth]{C55_M2O4.pdf}
%   \label{fig:DV_xx}
% \end{figure}

% \begin{figure}[t]
%   \centering
%   (a)\includegraphics[width=0.45\columnwidth]{C66_LMO.pdf}
%   (b)\includegraphics[width=0.45\columnwidth]{C66_NMO.pdf}\\
%   (c)\includegraphics[width=0.45\columnwidth]{C66_LM2O4.pdf}
%   (d)\includegraphics[width=0.45\columnwidth]{C66_MO.pdf}\\
%   (e)\includegraphics[width=0.45\columnwidth]{C66_M2O4.pdf}
%   \label{fig:DV_xx}
% \end{figure}

\begin{acknowledgments}
The authors thank the French National Research Agency (STORE-EX Labex Project ANR-10-LABX-76-01) for financial support.
\end{acknowledgments}

%\appendix
%\bibliographystyle{aipauth4-1}
%\bibliographystyle{h-physrev}
%\bibliography{Biblio}

%

\end{document}